\documentclass[doublecol,figures]{epl2}
\usepackage{amsmath}
\usepackage{wasysym}

\title{Polymer chain scission at constant tension - \\an example of
force-induced  collective behaviour}
\shorttitle{Scission of a polymer chain}

\author{
Jaroslaw Paturej\inst{1,2}\thanks{
E-mail: \email{jpaturej@univ.szczecin.pl}} \and
 Andrey Milchev\inst{1,3} \and
Vakhtang G. Rostiashvili\inst{1} \and Thomas A. Vilgis\inst{1}}
\shortauthor{J. Paturej \etal}

\institute{ \inst{1} Max Planck Institute for Polymer Research, 10 Ackermannweg,
55128 Mainz, Germany\\
\inst{2} Institute of Physics, University of Szczecin, Wielkopolska 15,
70451 Szczecin, Poland\\
\inst{3} Institute for Physical Chemistry, Bulgarian Academy of Sciences, 1113
Sofia, Bulgaria}


\abstract{The breakage of a polymer chain of segments, coupled by anharmonic
bonds with applied constant external tensile force is studied by means of
Molecular Dynamics simulation. We show that the mean life time of the chain
becomes progressively independent of the number of bonds as the pulling force
grows. The latter affects also the rupture rates of individual bonds along the
polymer backbone manifesting the essential role of inertial effects in the
fragmentation process. The role of local defects, temperature and friction  in
the scission kinetics is also examined.}

\begin{document}

\maketitle

\section{Introduction}

The understanding of a  great variety of phenomena related to stability,
fracture, and elastic behavior of materials requires fundamental knowledge of the
intermolecular dynamics of bond breakage.  In most cases scission of bonds may
be caused by mechanical load, irradiation, or just increase in temperature.
Examples related to this field are diverse and include mechanical fracture of
materials \cite{kausch,crist}, polymer rupture
\cite{garnier,grandbois,saitta,maroja,rohrig}, adhesion \cite{gersappe},
friction \cite{filippov}, mechanochemistry \cite{aktah,beyer}.  Recently, there
has been an enormous increase of interest in polymer fragmentation due to the
possibility of biomolecule's micromanipulation in experiments using force
spectroscopy methods \cite{harris,ober,greenleaf}.This has motivated
also theoretical investigations and computer experiments \cite{dudko}.

In particular, the problem of polymer fragmentation has got a longstanding
history in scientific literature. The treatment of bond rupture as a kinetic
process dates back to the publications of Bueche \cite{bueche} and Zhurkov \etal
\cite{zhurkov}. In the recent years these seminal papers  have been complemented
by a variety of computer experiments. Molecular Dynamics (MD) simulations of
chain rupture at constant stretching {\it strain} has been carried out, whereby
harmonic \cite{doerr,lee}, Morse \cite{stember,bolton,puthur} or Lennard-Jones
\cite{o1,o2,o3,o4} interactions have been employed. A theoretical interpretation
of MD results, based on an effectively one-particle model (Kramers rate theory)
has been suggested \cite{o3,o4}. On the other hand, an analytical treatment of a
polymer fragmentation under  {\it constant stress} have been proposed in terms
of many-particle version of transition state theory \cite{puthur}.

Recently, we proposed a description  of linear polymer scission under constant
tensile force by using the multidimensional Langer-Kramers theory, which was
found to compare favorably with the results of MD simulations \cite{gosh}.
Within this approach a single bond rupture is seen as a thermally activated
escape from the bottom of a potential well. The life time $\tau$ before a bond
scission takes place, is determined by diffusive crossing of an energy barrier
$E_b$ that is reduced  under the applied external force $f$. The adopted
theoretical treatment assumes a single collective unstable modes as being mainly
responsible for chain breakage. Such unstable mode peaks around an "endangered"
bond of negative spring constant and decays exponentially towards both chains
end. Similar collectivity effect has also been reported in the case of ring
polymers stretched with constant strain \cite{sain}.

In this letter we report some new results pertaining to the rupture kinetics of
single $1D$ and $3D$ polymer chains induced by constant tensile stress in a
broad interval of pulling forces.

\section{The model}
As in our previous work~\cite{gosh}, we use a coarse-grained model of
a polymer chain of $N$ beads connected by bonds, whereby each bond of
length $b$ is described by a Morse potential $V^{\mbox{\tiny M}} (r) = D \lbrace
1-\exp [ -a(r-b)] \rbrace^2$, with $a$ being a constant, $a=1$, that determines
bond elasticity,

The dissociation energy $D$ of a given bond is measured in units of $k_BT$,
where $k_B$ is the Boltzmann constant and $T$ denotes temperature. Since
$V^{\mbox{\tiny M}}(0) \approx 2$, the Morse potential is only weakly repulsive
and segments could partially penetrate one another at $r<b$. Therefore, in order
to allow properly for the excluded volume interactions between bonded particles,
we take the bond potential as a sum of $V^{\mbox{\tiny M}}$ and the so called
Weeks-Chandler-Anderson (WCA) potential $V^{\mbox{\tiny WCA}} = 4\epsilon\left[
(\frac\sigma r)^{12} - (\frac\sigma r)^6 + \frac 14
\right]\theta(2^{1/6}\sigma-r)$, with $\theta(x) = 0$ or 1 for $x<0$ or $x\geq
0$, and $\epsilon = 1$, $\sigma = 1$. The parameter $\sigma$ sets up the length
scale of equilibrium monomer size $b=2^{1/6}\sigma\approx 1.12$. The nonbonded
interactions between monomers are also taken into account by means of the WCA
potential.

The dynamics of the chain is obtain by solving a Langevin equation
for the position
$\mathbf q_n=[x_n,y_n,z_n]$ of each bead in the chain, 
$m\ddot{\mathbf q}_n = \mathbf F_n^{\mbox{\tiny M}} + \mathbf F_n^{\mbox{\tiny
WCA}} -\gamma\dot{\mathbf q}_n + \mathbf R_n(t) +\mathbf f_n\delta_{nN} \; 
(n,\ldots,N)$,
which describes the Brownian motion of a set of bonded particles whereby the
last of them is subjected to external stretching force $\mathbf f = [f,0,0]$.
The influence of solvent is split into slowly evolving viscous force and rapidly
fluctuating stochastic force. The random, Gaussian force $\mathbf R_n$ is
related to friction coefficient $\gamma$ by the fluctuation-dissipation theorem.
It should be noted that we consider only grafted chains, i.e.~one of the chain
ends is fixed in space. The integration step is $0.002$ time units (t.u.) and
time in measured in units of $\sqrt{m/\sigma^2D}$, where $m$ denotes the mass of
the beads, $m=1$.

We start the simulation with all beads placed at distance $b$ from each other,
and then we let the chain to equilibrate in the Langevin heat bath. Due to the
presence of the external pulling force, the equilibrium configuration of the
chain is more or less stretched and deviates markedly from coil shape. Once
equilibration is achieved, time is set to zero and one measures the elapsed time
$\tau$ before any of the bonds exceeds certain extension $r_h$, which sets the
criterion for considering such bond broken. We use a large value for the
critical bond extension, $r_h=5b$, which is defined as a threshold to a broken
state. This convention is based on our checks that the probability for
recombination of bonds, stretched beyond $r_h$, is vanishingly small. We repeat
this procedure for a large number of events $5\times 10^4$ so as to determine
the mean rupture time $\langle\tau\rangle$ which we refer as Mean First Breakage
Time (MFBT). The details of this method can be found in Ref.~\cite{gosh}.

\section{MD-results}
In our computer experiments we focused on the following most salient properties
of the bond breakage process:
 \subsection{Chain length dependence of the MFBT $\langle\tau\rangle$}
\begin{figure}
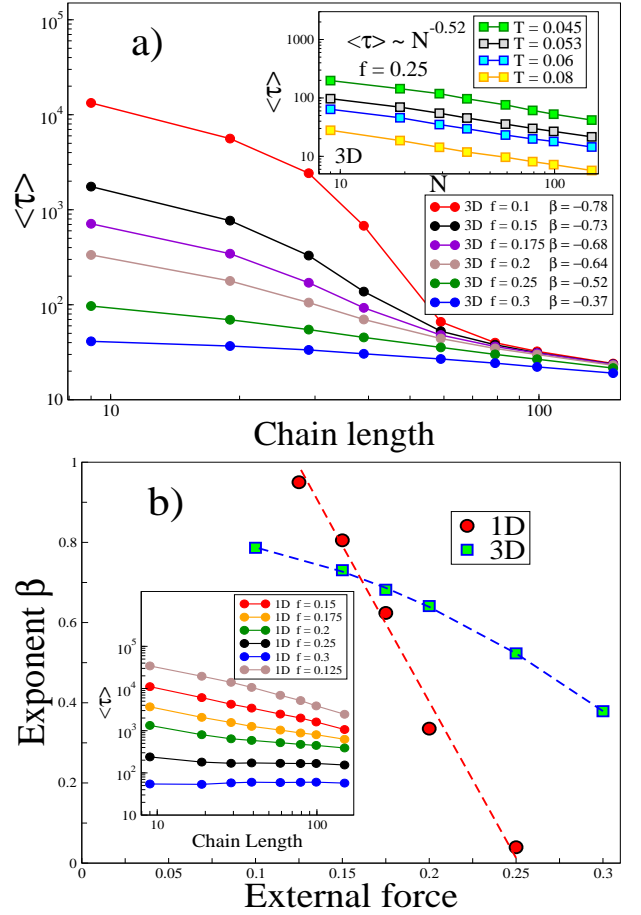

\onefigure[height=6.cm,width=8.cm]{3D_big_2_zero_force.eps}
\onefigure[height=6.cm,width=8.cm]{slopes_and_1D.eps}
\caption{a) Mean first breakage time $\langle \tau \rangle$ vs.~ $N$ for a $3D$
chain. In the legend slopes of fitting lines $\langle \tau \rangle \propto
N^{-\beta} $ are presented which were found in the range $N=80$--$150$. The
inset shows $\langle \tau \rangle$ against $N$ for $3D$ chains stretched by
force $f=0.25$ at different temperatures. b) Variation of slope $\beta$ with
external pulling force $f$ for chains in $1D$ and $3D$. The inset shows
$\langle \tau \rangle$ vs.~$N$ for a $1D$ chain. Parameters of the heat bath are
temperature $T=0.53$ and friction $\gamma=0.25.$}
\label{cld}
\end{figure}

In Fig.~\ref{cld}a) and the inset of Fig.~\ref{cld}a) we present numerical
results for $\langle \tau \rangle$ as a function of the number of beads $N$ for
chains that are stretched in the interval $0.1\geq f\geq 0.3$ of pulling forces
both in $1D$ and $3D$. Regardless of dimensionality of the examined systems,
for a given value of $f$ one observes a power-law decrease, $\langle \tau
\rangle \propto N^{-\beta}$. This relationship is found for sufficiently long
chains (asymptotic limit) -- $N \apprge 80$, where finite-size effects do not
play a role. Additionally, from the inset of Fig.~\ref{cld}a) we observe that
there is no impact of the temperature on the value of the slope $\beta$.
Furthermore, Fig.~\ref{cld}b) indicates that  with growing tensile strength the
life time $\langle \tau \rangle$ becomes nearly independent of $N$ which is
among the most important results of this study. This independence is fully
consistent with our recent findings \cite{gosh} for relatively strong pulling
force, $f = 0.25$. It was shown in \cite{gosh} that in this case the process of
bond scission is governed by a {\it collective} unstable mode peaked around an
''endangered'' bond (i.e.~a bond with negative spring constant) and decays
exponentially towards both chain's ends. On the other hand, in the opposite
limit of thermal degradation of  polymers ($f=0$) (i.e., in the so-called {\em
thermolysis}) the total probability for scission of a polymer with $N$ bonds
within a certain time interval is $N$ times larger than that for a single bond
which is what one would expect if bonds do break entirely at random and
independent of one another. The latter leads to the relationship $\langle \tau
\rangle \propto N^{-1}$ which has been seen recently in computer simulations of
harmonic \cite{fugmann} and anharmonic \cite{paturej} polymer chain models.
Moreover, Fig.~\ref{cld}b) clearly shows that with increasing pulling force $f$
the exponent $\beta$ gradually decreases within the interval $0 < \beta < 1$.
Thus the slope $\beta$ can be treated as a quantitative measure of the degree
of cooperativity in rupture events . As the slope $\beta$ decreases, the nature
of scission events become more and more collective.

It is pertinent to note that in the literature one finds conflicting data
regarding the $\langle \tau \rangle$ vs.~$N$ dependence. For example, in
\cite{o4} it is claimed that $\langle \tau \rangle \propto N^{-1}$, but a more
close inspection of the Fig. 4 in this work where the dependence is shown gives
for the exponent $\beta = 0.17 \div 0.2$. In another work \cite{sain}, a chain
rupture under fixed strain has been investigated by making use the many-body
Langer-Kramers theory and MD-simulation. On the one hand, the authors discuss
the presence of collective unstable mode (cf. \cite{gosh}) but on the other hand
they claim that for the total chain again $\langle \tau \rangle \propto N^{-1}$
which should be considered as a hallmark of missing collectivity. Unfortunately,
there have been no corresponding simulations which would have confirmed the
conclusion made in \cite{sain}.

\subsection{Dependence of MFBT on Pulling Force}
\begin{figure}[htb]
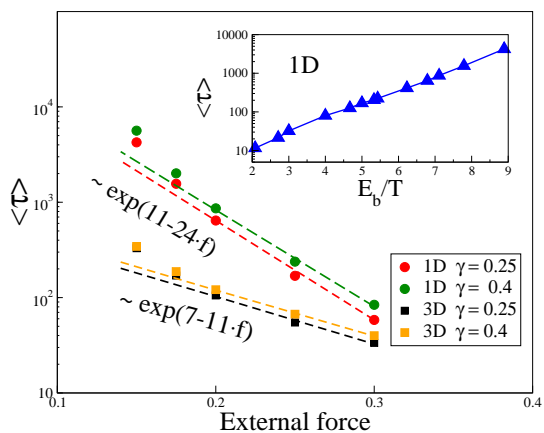

\onefigure[scale=0.28]{tauvsf3d.eps}
\caption{Force-dependent mean first breakage time
for a $1D$ and $3D$ chains with $N=30$.
The inset shows of  $\langle\tau\rangle$ vs.~ $E_b/T$ for $1D$ system with
$T=0.053$ and $\gamma = 0.25$.}
\label{extforce}
\end{figure}

The dependence of the MFBT $\langle\tau\rangle$ on external force $f$ for $1D$
and $3D$ chains composed of $N=30$ beads is shown in Fig.~\ref{extforce}.
Evidently for sufficiently strong stretching forces $f \apprge 0.175$ an
exponential decay $\langle \tau \rangle \propto e^{(E_0 -\alpha f)/k_B T}$  is
observed. The main reason of this is the following: As the pulling force grows,
the  energy barrier, which separates intact bonds from the broken ones,
declines. As a consequence, $\langle \tau \rangle$ decreases. One should  note
that the parameters $\alpha$ and $E_0$ change only slightly with the coupling
parameter $\gamma$ of the thermostat.
Fig.~\ref{extforce} indicates also a considerable difference in the values of
$\alpha$  between $1D$ and $3D$. In the inset of Fig.~\ref{extforce} we present
$\langle \tau \rangle$ as a function of the ratio $E_b/T$ of the barrier height
to temperature. This finding is in agreement with the understanding of the
polymer rupture as a thermally activated process \cite{bueche,zhurkov} and is
manifested by an Arrhenian relationship -- $\langle \tau \rangle \propto
e^{E_b/T}$, where $E_b = E_0 - \alpha f$.

\subsection{Life-time probability distribution $W(t)$}
\begin{figure}[htb]
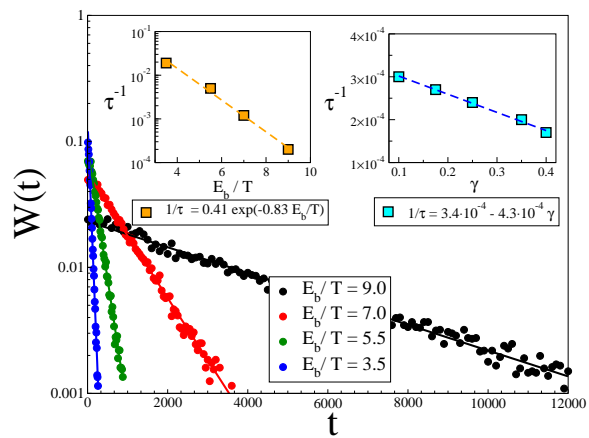

\onefigure[scale=0.28]{wtau_E_b_T.eps}
\caption{Life-time probability distributions $W(t)$ for different height of the
energy barrier $E_b/T$ in $1D$. Here the chain length is  $N=30$, the pulling
force $f=0.15$, and $\gamma = 0.25$. Symbols denote simulation results and full
lines stand for fitting functions $W(t) \propto \exp(-t/\tau)$. The two insets
show the dependence of  $\tau^{-1}$ on $E_b/T$ [left panel], and on $\gamma$
[right panel].}
\label{W}
\end{figure}

In Fig.~\ref{W} we display the probability distribution function $W(t)$ of the
observed scission times $t$ for several ratios $E_b/T$ of the barrier height to
temperature  in the case of $1D$ chain composed of $N=30$ beads. It appears that
$W(t)$ goes asymptotically as $W(t) \propto e^{-t/\tau(E_b/T)}$ in accordance
with our recent findings \cite{gosh}. Additionally from the inset [left panel]
of Fig.~\ref{W} one may easily verify that the characteristic time goes as
$\tau(E_b/T) \propto \exp(E_b/T)$. Moreover,  a simple linear relationship has
been found for friction dependence of $\tau(\gamma)$ as shown in the inset
[right panel] of Fig.~\ref{W}.

\subsection{Rupture probability histograms}
\begin{figure}
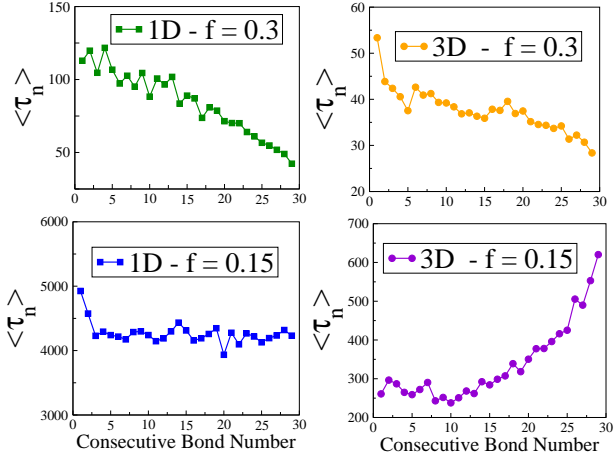

\onefigure[height=6.cm,width=8.cm]{tau_n_N.eps}
\caption{Variation of the mean life time $\langle \tau_n \rangle$ with
consecutive bond number for a $30$-particle chain in one and in three
dimensions. The polymer is stretched by a force $f=0.15$ or $f= 0.3$ at $T=0.053$
and $\gamma=0.25$.}
\label{mtvsc}
\end{figure}

In Fig.~\ref{mtvsc} we show the MFBT $\langle\tau_n\rangle$ of the {\em
individual} bonds for chains composed of $N=30$ beads in $1D$ and $3D$. We
compare the results for different pulling forces as indicated in the legend. For
the case $f=0.3$ one may readily verify that  the  bonds located in the vicinity
of the grafted bead live nearly twice as long as those close to the loose end
where the tensile force is applied,  regardless of dimensionality.
In contrast, chains stretched by a gentle force, $f=0.15$, display very
different distributions of $\langle \tau_n \rangle$. In the case of a $3D$
system the bonds at the tethered end live on the average significantly ($2\div3$
times) shorter than those close to pulled end. Evidently, staring from $n
\apprge 10$ the lifetime of bonds progressively increases with increasing
proximity to the free chain end. In the case  of $1D$  string stretched with
$f=0.15$ this effect is missing and $\langle \tau_n \rangle$ is uniformly
distributed along the polymer backbone apart from the first terminal bond which
lives $\approx 20\%$ longer. As expected, all distributions displayed in
Fig.~\ref{mtvsc} are asymmetric due to the constraint imposed on the motion of
the first bead in the chain.
\begin{figure}[bhpt]
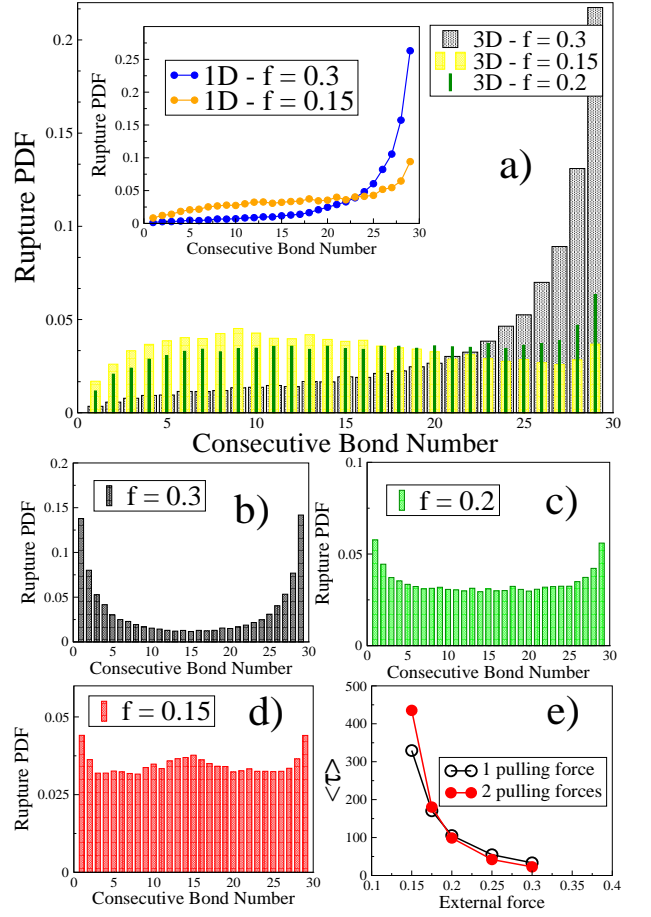

\onefigure[height=6.cm,width=8.cm]{rupPDF_1Dvs3D.eps}
\onefigure[height=6.cm,width=8.cm]{2pull_forces.eps}
\caption{a) Rupture probability histograms for $1D$ and $3D$
 chains composed of $N=30$ for different pulling forces
 as indicated.
b)--d)
Rupture probability histograms for $3D$ chains made of
$30$ particles pulled at {\em both} ends with a tensile force $f$.
e) Variation of $\langle \tau \rangle$ with force $f$  for chains pulled either
at the  one end, or at both ends. Here $T=0.053$ and $\gamma = 0.25$.}
\label{ruphist}
\end{figure}

In Fig.~\ref{ruphist} we present the probability for bond scission of individual
bonds in $1D$ and $3D$ for several strengths of the pulling force. The
histograms display the (normalized) rate at which a certain bond $n$ along the
polymer backbone breaks. From the inspection of  Fig.~\ref{ruphist}a) one sees
that the preferential scission of the bonds with particular consecutive bond
number essentially depends on the value of force. For strong pulling $f=0.3$ one
finds that the terminal bond which is subjected to pulling as well as the bonds
in its neighborhood break more frequently than whose around fixed end.
Evidently,  in this case the rupture rate decreases steadily from the free chain
end to the tethered one. A similar scission scenario is visible also for the
$1D$ chain as shown in the inset of Fig.~\ref{ruphist}a).

In contrast, as the stretching force is decreased, the corresponding rupture
histogram for a $3D$ chain becomes flatter. For $f=0.2$ 
the distribution of
 scission rates becomes uniform exept for the bonds in the vicinity of both ends. A
further decrease of the pulling force results in a qualitative change in the
distribution. Evidently, for $f=0.15$ the bonds in the middle of the chain,
which are also somewhat closer to the fixed chain end, become more vulnerable as
compared to those at the chain ends. Note that for the smallest pulling force
($f\apprle 0.15$) the rupture histogram already resembles the respective
histogram in the case of thermal degradation of a polymer \cite{paturej} which
takes place in the absence of externally induced tension. If two pulling forces
are applied simultaneously to both chain ends, Fig.~\ref{ruphist}b)-d), one finds
expectedly a symmetric scission probability distribution regarding bond number,
even though $\langle \tau \rangle$ hardly changes - Fig.~\ref{ruphist}e). Again.
the scission rate of the terminal bonds goes down with decreasing strength of
the pulling force $f$, and for $f = 0.15$ a local maximum in the rate builds up
for the bonds that are in the middle of the chain.

\begin{figure}[bhpt]
\onefigure[scale=0.3,angle=270]{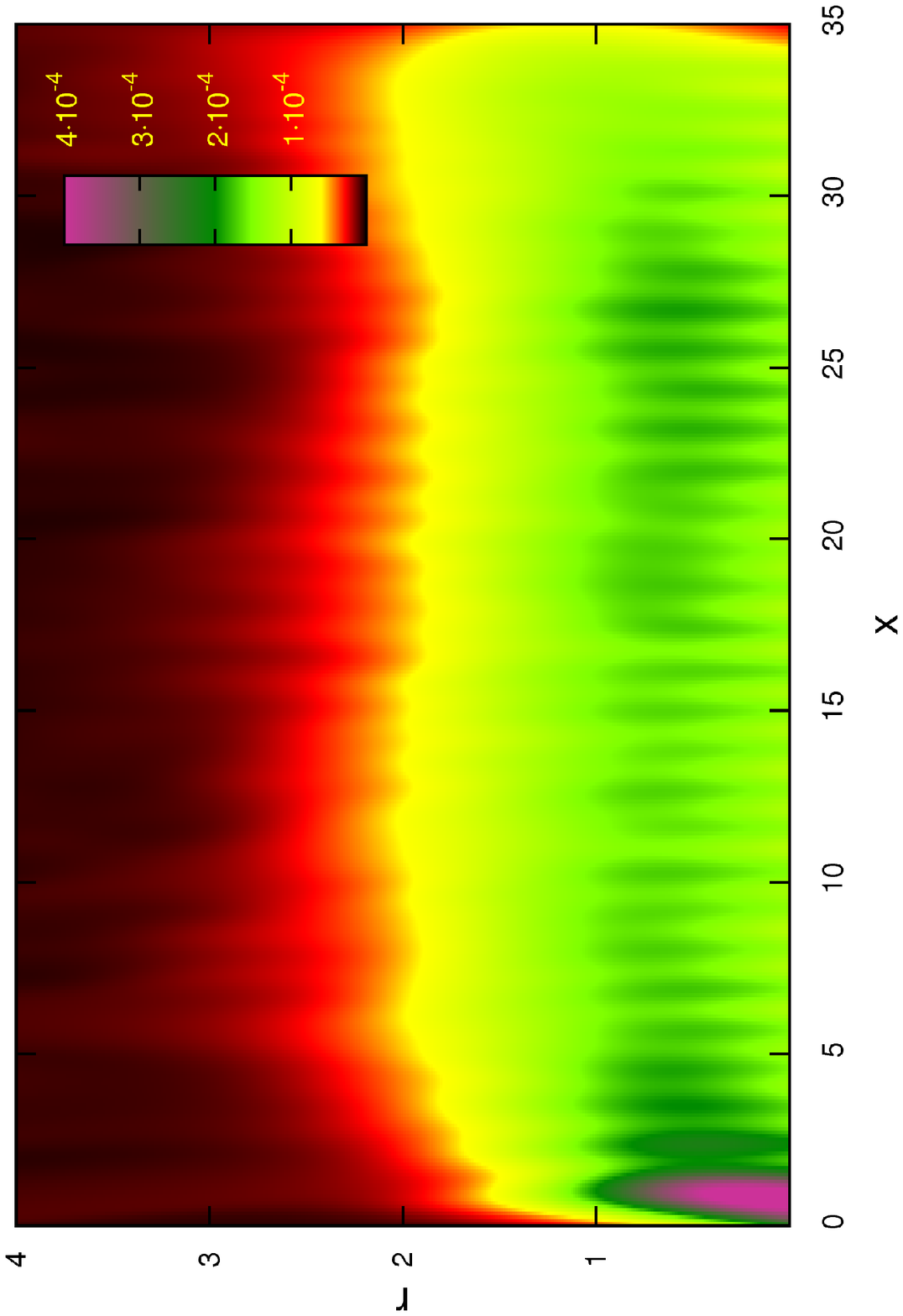}
\vspace*{-0.5cm}
\onefigure[scale=0.3,angle=270]{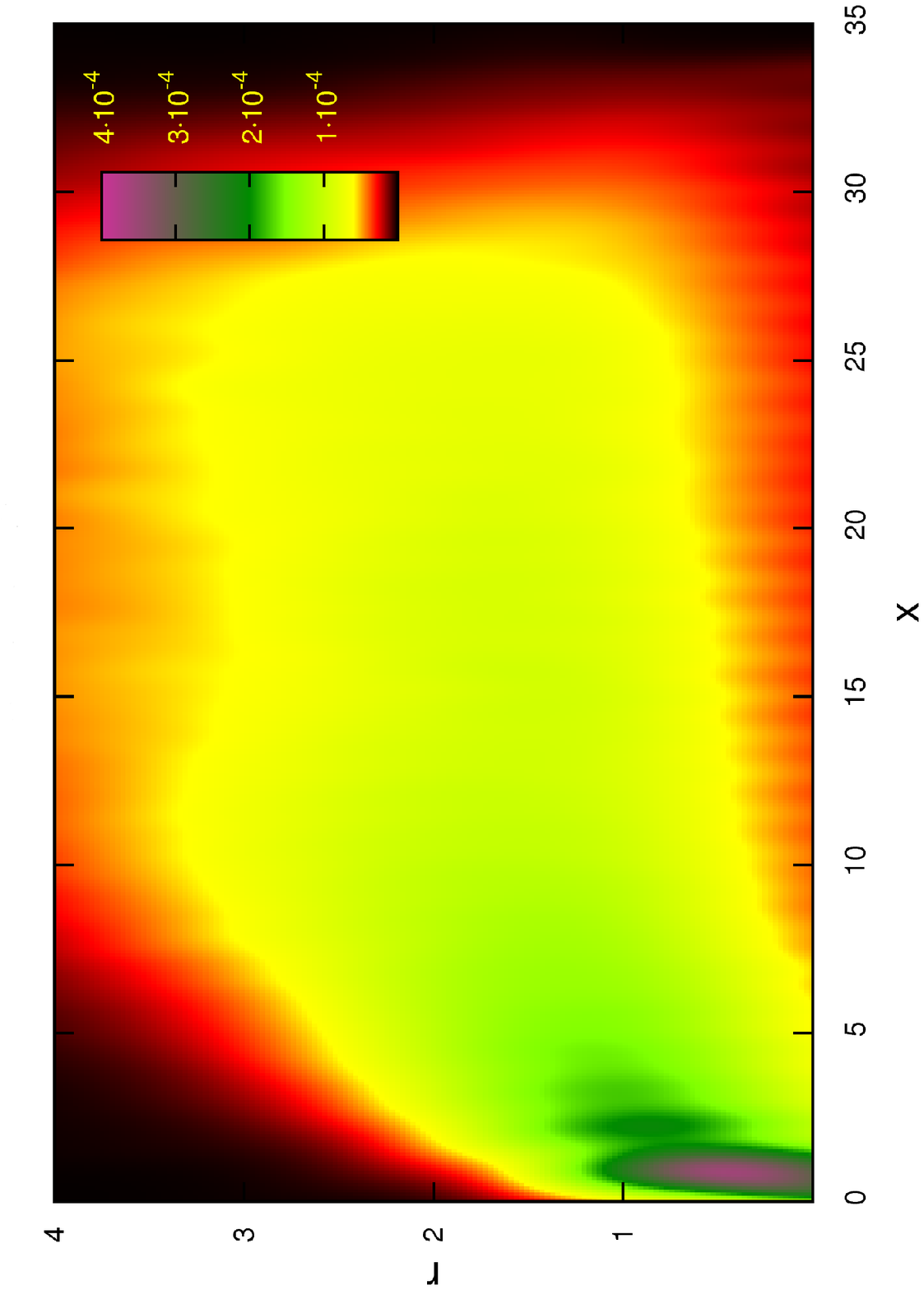}
\caption{Probability density distribution $P(x,r)$ of beads  in a $3D$ chain
with $N=30$ particles at force: $f=0.3$ [upper panel], and
$f=0.15$ [lower panel]. The $x$-axis coincides with the direction of pulling
force whereas $r=\sqrt{y^2+z^2}$ denotes radial component of the bead position.
Different colors indicate the value of the PDF as indicated in the legend. Here
$T=0.053$.}
\label{cigar}
\end{figure}

How can such an inhomogeneity in the probability of bond rupture be understood?
A possible explanation of the change in the location of preferential breakdown
sites along  the chain may be gained by Fig.~\ref{cigar}. In this figure we
present maps of the density distribution $P(x,r)$ of bead positions where $x$
is measured in direction of the pulling force $f$ whereas $r=\sqrt{y^2+z^2}$
denotes the radial component. Fig.~\ref{cigar}~[upper panel] indicates that
at high stretching ($f=0.3$) the most probable position of the beads is along
the direction of the tensile force.
For $f=0.3$ (strong stretching) the chain conformation corresponds to a
quasi-$1D$ structure, and the transversal fluctuations are reduced.
In contrast, when the pulling force is weak ($f=0.15$) one finds from
Fig.~\ref{cigar}~[lower panel] that the individual beads are free to make big
excursions in space -- $P(x,r)$ is roughly two times  broader in the middle.
Thus, Fig.~\ref{cigar} suggests that the density maps comply with the rupture
histograms given in Fig.~\ref{ruphist}a). For $f=0.3$, due to larger freedom
around the pulled end, the end bonds break more easily in the terminal part of
the chain. When the force is weak, $f=0.15$, the beads become more mobile around
the center of the polymer which in turn leads to increased bonds scission rate
there.

\subsection{Chain Defects}
\begin{figure}[bhpt]
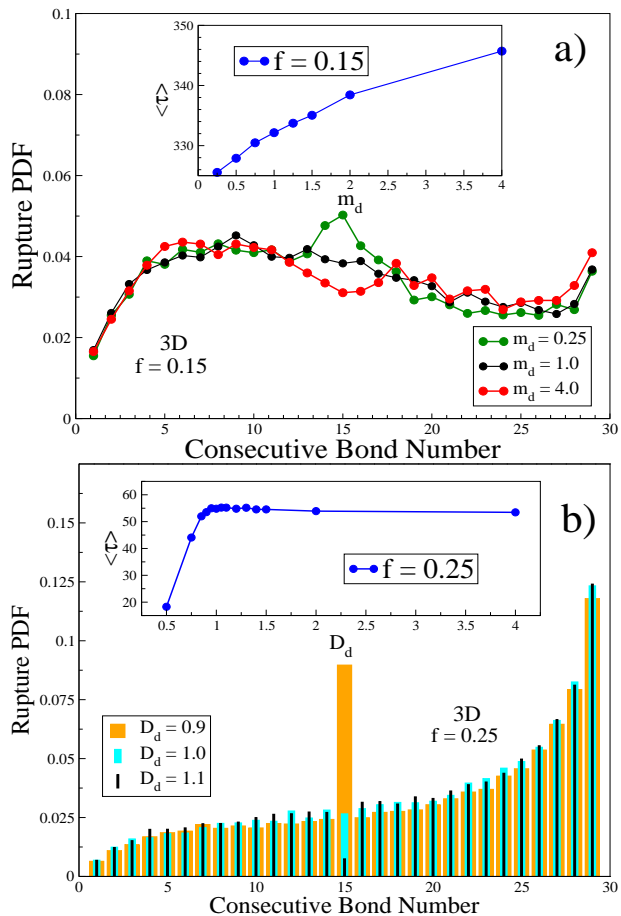

\onefigure[height=6.cm,width=8.cm]{massboxes.eps}
\onefigure[height=6.cm,width=8.cm]{Dboxesf_25.eps}
\caption{a) Rupture probability histograms for $30$-bead chains with a single
defect bead of  mass $m_d$ introduced in the center of a chain. Here $f=0.15$.
The inset shows the mean life time time dependence on the mass of the defect. b)
Rupture probability  histogram for chains with a single bond defect  ({\em bond
strength} $D_d$) in the middle of the chain. The inset shows dependence of mean
first breakage time on the strength of the defected bond. Parameters of the heat
bath are: $T=0.05328$ and $\gamma = 0.25$. }
\label{defect}
\end{figure}

In Fig.~\ref{defect} we present the results  of computer experiments concerning
rupture of $30$-particle chains in which a single defect is introduced. We focus
on two kind of defects. First we examine the effect of an "isotope"-like defect
in which a mass $m_d$ of the monomer located in the center  of a chain is
changed, while  the masses of all remaining beads remain unchanged and equal to
$m=1$. In Fig.~\ref{defect}a) we compare the rupture histogram for a chains with
such a heavy/light bead to the respective scission probability distribution in
a uniform system.
If the central bead is replaced by a particle which is lighter/heavier than the
rest of the segments, the rupture probability  increases/decreases in the
immediate vicinity of the introduced mass-defect. This effects a small group of
beads $\sim 3$ and is pronounced only when the stretching force is sufficiently
small ($f \apprle 0.2$). Evidently, a lighter particle can be kicked more easily
by the thermostat whereby inertial effects would stretch both adjacent bonds
beyond the scission threshold. In contrast, as is seen from the inset of
Fig.~\ref{defect}a), the MFBT $\langle \tau \rangle$ grows as the defected bead
becomes heavier. A defect with a larger mass is hard to accelerate, its mobility
is low, and its bonds remain unstretched. In some interval of time such defect
experiences many kicks which effectively  cancel each other before a dangerous
bond stretching occurs. Thus, the chain becomes locally more immune against
breakage events  which is reflected by the decrease in the  probability of
rupture.

As expected, the tensile strength of a particular bond (measured in units of $D$)
affects the scission process too. To see this we varied the bond strength $D_d$
of the middle bond. In Fig.~\ref{defect}b) we present rupture histograms for
chains with $N=30$ and different values of $D_d$. Clearly, the small variation
of bond strength ($\pm 10\%$) results in a change of the scission probability
which is located exactly at the defect position. In the inset of
Fig.~\ref{defect}b) we present the MFBT $\langle \tau \rangle$ as a function of
$D_d$. One can see that the introduction of weaker bonds $D_d<1$ in the chain
results in a decrease of  $\langle \tau \rangle$. This is due to the fact that
the lifetime $\tau_D$ for the weak bond is very short and therefore dominates
the mean $\langle \tau \rangle$. On the contrary, it appears that $\langle \tau
\rangle$ is not sensitive for defects with $D_d>1$. In this case defected bonds
appear to resist scission events which then happen predominantly in the
remaining bonds.

\section{Concluding remarks}
Our findings can be summarized as follows:
\begin{itemize}
\item The mean life time of the polymer chain at constant tensile force depends
on chain length like $\langle \tau \rangle \propto N^{-\beta}$ whereby the
power law exponent $\beta$ varies in the interval $0 < \beta < 1$. Generally,
it appears  that the exponent $\beta$ systematically declines as the external
pulling force $f$ grows. This behaviour indicates a growing degree of
cooperativity during the chain breakage as the pulling force $f$ is increased.

\item  The MFBT follows an Arhenian law $\langle \tau \rangle \propto \exp
(E_b/T)$ whereby the effective activation barrier for scission changes with
varying pulling force $f$ as $E_b = E_0 - \alpha f$ in line with earlier
theoretical predictions \cite{zhurkov}.
The scission times in a polymer chain under tension are exponentially
distributed, $W(t) \propto \exp(- t / \tau (E_b/T))$.

\item The rates of bond rupture are distributed differently along the polymer
backbone in the $1D$ and $3D$ chain models. In a $1D$ chain the rupture rate
steadily grows as one approaches the free chain end where the external pulling
force is applied whereas in a $3D$ chain bonds break predominantly in the
middle of the chain. Bond rupture histograms correlate with the degree
of spreading in the monomer density distribution, indicating that scissions
occur most frequently in those parts of the macromolecule which undergo large
fluctuations in position. Inertial effects and bead mobility provide a
plausible interpretation of the observed complexity of fragmentation kinetics.

\item The probability histograms for bond scission provide a clear picture of
the impact of polymer defects  on the fragmentation process and underline
thereby the role of  inertial effects. Bonds, connected to lighter segments
break become a preferred site of rupture whereas heavier segment stabilize the
chain.

\end{itemize}

\acknowledgments

J. P. would like to thank Michael Rubinstein for fruitful discussions. A. M. gratefully acknowledges support
by the Max-Planck-Institute for Polymer Research during the time of this investigation. This research has been
supported by the Deutsche Forschungsgemeinschaft (DFG), Grants SFB 625/B4 and FOR 597.

\end{document}